\documentclass[aps,prl,showpacs,amsmath,amssymb,superscriptaddress,twocolumn]{revtex4-1}

\usepackage{graphicx}
\usepackage{amsmath}
\usepackage{hyperref}
\usepackage{dcolumn}
\usepackage{bm}
\usepackage{color}
\usepackage{nicefrac}

\begin{document}

\title{Locating the QCD critical endpoint through finite-size scaling}

\author{N.~G.~Antoniou}
 \email[]{nantonio@phys.uoa.gr}
\affiliation{Faculty of Physics, University of Athens, GR-15784 Athens, Greece}

\author{F.~K.~Diakonos}
 \email[]{fdiakono@phys.uoa.gr}
\affiliation{Faculty of Physics, University of Athens, GR-15784 Athens, Greece}

\author{X.~N.~Maintas}
 \email[]{xmaintas@phys.uoa.gr}
\affiliation{Faculty of Physics, University of Athens, GR-15784 Athens, Greece}

\author{C.~E.~Tsagkarakis}
 \email[]{ctsagkarakis@phys.uoa.gr}
\affiliation{Faculty of Physics, University of Athens, GR-15784 Athens, Greece}

\date{\today}

\begin{abstract}
\noindent
Considering the 3d Ising universality class of the QCD critical endpoint we use a universal effective action for the description of the baryon-number density fluctuations around the critical region. Calculating the baryon-number multiplicity moments and determining their scaling with system's size we show that the critical region is very narrow in the direction of the baryon chemical potential $\mu$ and wide in the temperature direction $T$ for $T > T_c$. In this context, published experimental results on local proton density-fluctuation measurements obtained by intermittency analysis in transverse momentum space 
in NA49 central A+A collisions at $\sqrt{s_{NN}}=17.2$ GeV (A=C,Si,Pb), 
restrict significantly the location $(\mu_c,T_c)$ of the QCD critical endpoint. The main constraint is provided by the freeze-out chemical potential of the Si+Si system, which shows non-conventional baryon density fluctuations, restricting $(\mu_c,T_c)$ within a narrow domain, $119~\textrm{MeV} \leq T_c \leq 162~\textrm{MeV}$, $252~\textrm{MeV} \leq \mu_c \leq 258~\textrm{MeV}$, of the phase diagram.


\end{abstract}


\maketitle
The search for the QCD critical endpoint (CEP), remnant of the chiral symmetry breaking, at finite baryon density and high temperature, is the main task in contemporary relativistic ion collision experiments \cite{NA61,RHIC}. Fluctuation analysis with global \cite{Stephanov2005,Luo2017} and local measures \cite{Antoniou2006} is the basic tool to achieve this goal. Up to now, indication of such non-conventional fluctuations, which can be related to the CEP, has been observed in the freeze-out state of Si+Si central collisions at NA49 SPS experiment with beam energy $\sqrt{s_{NN}}=17.2$ GeV \cite{Anticic2015}. However, the strong background and the poor statistics in the corresponding data set, did not allow for convincing statements concerning the existence and the location of the CEP. Similarly, in RHIC BES-I program, a non-monotonic behaviour of $\kappa \sigma^2$ (kurtosis times the variance) for net-proton distribution, compatible with theoretical proposals \cite{Stephanov2011}, was observed \cite{Luo2016,Luo2017} but a conclusive evidence for the location of the critical point is still pending, so its experimental hunt continues. From the theoretical side the efforts are focused on Lattice QCD calculations at finite chemical potential in order to obtain the QCD phase diagram from first principles and predict the location of the CEP. Unfortunately, the until now obtained Lattice results depend strongly on the method used to handle the well known sign problem and they do not converge to a well defined critical chemical potential value \cite{Katz2004,Gupta2008}. Therefore a first principle prediction of the QCD CEP location, the holly grail of the physics of strongly interacting matter in our times, is still missing. 

In the present Letter we will make an effort to estimate the QCD CEP location employing an appropriate effective action for the thermodynamic description of the baryonic fluid around the critical region. To this end, we will assume that CEP belongs to the 3d Ising universality class, a hypothesis which is strongly supported by several theoretical works \cite{Gavin1994,Stephanov1998,Halasz1998,Berges1999,Karsch2001}. In this context, a universal effective action, found on the basis of a Monte-Carlo simulation of the 3d Ising system in an external field \cite{Tsypin1994}, is an appropriate tool for the formulation of the QCD critical properties. Introducing a dimensionless scalar field $\phi=\beta_c^3 n_b$ (order parameter) with $\beta_c=1/k_B T_c$ and $n_b$ the baryon-number density, the effective action is written as follows:
\begin{eqnarray}
S_{eff}&=&\int_V d^3 \hat{\mathbf{x}} \left[ \frac{1}{2} \vert \hat{\nabla} \phi \vert^2 + U(\phi) - \hat{h} \phi \right]~~~;~~~T \geq T_c \nonumber \\
U(\phi)&=&\frac{1}{2} \hat{m}^2 \phi^2 + \hat{m} g_4 \phi^4 + g_6 \phi^6
\label{eq:1}
\end{eqnarray}
In Eq.~(\ref{eq:1}) the variables with a "hat" are dimensionless: $\hat{x}_i=x_i \beta_c^{-1}$, $\hat{m}=\beta_c m$ ($m=\xi^{-1}$, $\xi$ being the correlation length), $\hat{h}=(\mu-\mu_c) \beta_c$ (ordering field) and $g_4=0.97 \pm 0.02$, $g_6=2.05 \pm 0.15$ are universal dimensionless couplings \cite{Tsypin1994}. 

The partition function, on the basis of Eq.~(\ref{eq:1}) is written schematically:
\begin{equation}
\mathcal{Z}=\int \left[ \mathcal{D} \phi \right] \exp(-S_{eff})
\label{eq:2}
\end{equation}
and considering the ensemble of constant $\phi$-configurations ($\hat{\nabla} \phi =0$) we obtain a grand-canonical expansion:
\begin{equation}
\mathcal{Z}=\displaystyle{\sum_{N=0}^M} \zeta^N \exp\left[-\frac{1}{2} \hat{m}^2 \frac{N^2}{M}-g_4 \hat{m} \frac{N^4}{M^3} - g_6 \frac{N^6}{M^5}\right]
\label{eq:3}
\end{equation}
where $\zeta=\exp\left(\frac{\mu-\mu_c}{k_B T_c}\right)$, $M=\frac{V}{\beta_c^3}$, $\hat{m} = \beta_c \xi^{-1}=\left[\frac{T-T_c}{T_c}\right]^{\nu}$ and $\nu \approx 2/3$ for the 3d Ising universality class \cite{Pelissetto2002}. Our aim is to calculate the 
baryon-number distribution moments:
\begin{equation}
\langle N^k \rangle=\displaystyle{\frac{1}{\mathcal{Z}}\sum_{N=0}^M} N^k \zeta^N \exp\left[-\frac{1}{2} \hat{m}^2 \frac{N^2}{M} -g_4 \hat{m} \frac{N^4}{M^3} - g_6 \frac{N^6}{M^5}\right]
\label{eq:4}
\end{equation}
with $k=1,2,..$ and explore their scaling behaviour with the system's size $M$ around the critical region. At the critical point $\zeta_c=1$ ($\mu=\mu_c$), $\hat{m}_c=0$ ($T=T_c$) these moments obey the scaling law:
\begin{equation}    
\langle N^k \rangle \sim M^{k q}~~~~;~~~~q=d_F/d,~k=1,2,..
\label{eq:5}
\end{equation}
where $d$ is the embedding dimension of the considered system and $d_F$ the fractal dimension related to the critical fluctuations of the baryon density \cite{Antoniou2017}.

Our strategy to determine the location of the QCD CEP is the following. First we will estimate the size of the critical region based on the scaling behaviour of $\langle N \rangle$ with the system's size $M$. Then, using the published NA49 results on proton intermittency analysis in central A+A collisions at $\sqrt{s_{NN}}=17.2$ GeV\cite{Anticic2015} we will constrain the location of the CEP. For the second step it is crucial that the critical exponent $q$ in Eq.~(\ref{eq:5}) is directly related to the intermittency index $\phi_2$ measured in the proton intermittency analysis. Let us start with the estimation of the critical region. We determine the dependence of $\langle N \rangle$ on $M$ for different values of $\zeta$ and $\hat{m}$. Exactly at the critical point $(\zeta_c,\hat{m}_c)=(1,0)$ the critical exponent $q$ attains the value $5/6$ for the 3d-Ising universality class ($\delta=5$) extracted from the representation (\ref{eq:3}). A direct calculation of $\langle N \rangle$ as a function of $M$ from the partition function in Eq.~(\ref{eq:3}) shows that, departing slightly from the critical point leads to a behaviour $\langle N \rangle \sim M^{\tilde{q}}$ for $M \gg 1$ with $\tilde{q} \neq q$. Outside the critical region $\tilde{q}=1$. Varying $\hat{m}$ and $\zeta$ we  may also enter to the $\phi^4$-dominance region when the last term in the effective action (\ref{eq:1}) becomes suppressed with respect to the other two effective potential terms. In that case $\tilde{q}=3/4$ (mean field universality class, $\delta=3$) and the information of the 3d-Ising critical exponent $q$ is again lost. Thus, we consider as critical region of the QCD CEP the domain in the $(\zeta,\hat{m})$-plane for which 
\begin{equation}
\langle N \rangle \sim M^{\tilde{q}}~~~~;~~~~3/4 < \tilde{q} < 1
\label{eq:6}
\end{equation}
holds. Since the critical region depends in general on the size $M$, a supplementary constraint that the correlation length is greater than the linear size of the system, compatible with finite-size scaling theory \cite{Stinchcombe1988,Ortmanns1996}, is certainly needed. To keep contact with $M$-values realistic for the size of the fireball produced in relativistic ion collisions we explore the validity of the scaling law (\ref{eq:6}) for $20 < M < 700$. This estimated range of $M$-values contains all sizes of nuclei ranging from Be ($R_{\textrm{Be}} \approx 2.6~\textrm{fm}$) to Pb ($R_{\textrm{Pb}} \approx 7.5~\textrm{fm}$), assuming $T_c \approx 150$ MeV. In Fig.~1a the red shaded area denotes the critical region, i.e. the domain in $(\zeta,\hat{m})$-plane for which the above constraints hold. We observe that the critical region is quite extended in the upper $\hat{m}$-direction but it is very narrow in the $\zeta$-direction. This is a crucial property restricting the location of the CEP. The blue line in Fig.~1a is the location of the $(\zeta,\hat{m})$ pairs which lead to a scaling of $\langle N \rangle$ with $\tilde{q}=0.96$. This is the mean value of the intermittency index $\phi_2$ found in the SPS NA49-data analysis of the proton-density fluctuations in transverse momentum space for  Si+Si central collisions at $\sqrt{s_{NN}}=17.2$ GeV. It can be shown that the intermittency index $\phi_2$ in transverse momentum space is equal to the exponent $\tilde{q}$ in 3d configuration space \cite{Antoniou2006,Antoniou2016}. Thus, the Si+Si freeze-out state should lie on this blue line. Taking the freeze-out data of the fireball created in central collisions of Si+Si at $\sqrt{s_{NN}}=17.2$ GeV to be $(\mu,T)$=$(260,162)$ MeV, as reported in \cite{Becattini2006}, the condition of lying on the blue line provides a relation between $\mu_c$ and $T_c$. This relation, determining all the possible freeze-out values for the QCD CEP compatible with our analysis and with the intermittency results in \cite{Anticic2015}, is presented graphically in Fig.~1b. The freeze-out temperature of Si+Si sets an upper bound on the critical temperature $T_c < 162$ MeV. A lower bound on $T_c$ is provided by the requirement that the correlation length $\xi$ is greater than the linear system size for the occurrence of critical fluctuations as stated above. For the smallest considered system, Be, with a radius of $\approx 2.6$ fm, we obtain the upper bound $\frac{T-T_c}{T_c} < 0.36$ which covers also the case of Si and leads (Fig.~1b) to the lower bound on the critical temperature $T_c > 119~\textrm{MeV}$. Thus, in the plot of Fig.~1b we show only the allowed range $119~\textrm{MeV} < T_c < 162~\textrm{MeV}$. We observe that the corresponding critical chemical potential domain is very narrow: $252~\textrm{MeV} < \mu_c < 258~\textrm{MeV}$. In fact, using the available Lattice-QCD estimates of the critical temperature $T_c \approx 146$ MeV \cite{Gupta2008} and $T_c \approx 153$ MeV \cite{Katz2004} as the boarders of a critical zone (red shaded domain in Fig.~1b) we obtain a very narrow range for the critical chemical potential $256~\textrm{MeV} < \mu_c < 257~\textrm{MeV}$. For completeness we add in Fig.~1a the freeze-out states for the central collisions of the other two systems, Pb+Pb and C+C, considered in the NA49 experiment (at $\sqrt{s_{NN}}=17.2$ GeV). We clearly observe that these systems lie outside the critical (shaded) region although their freeze-out chemical-potential values do not differ so much. The reason is {\em the narrowness of the critical region in the chemical potential direction} which possesses a linear size of $5$ MeV for $T=T_c$ and $T_c \approx 150$ MeV.

\begin{figure}[tbp]
\centering
\includegraphics[width=0.55\textwidth]{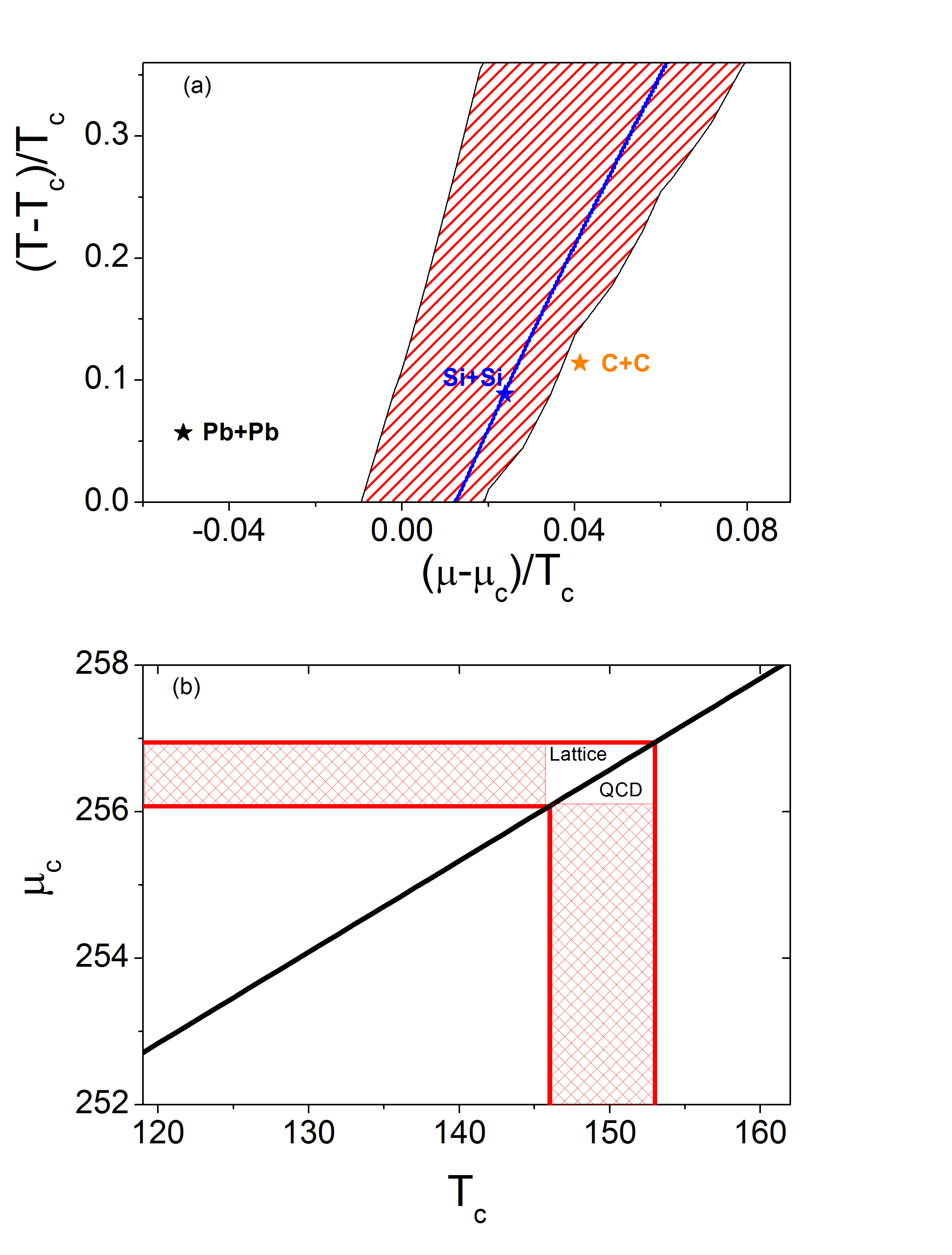}
\caption{(a) The critical region (red shaded area) in the plane $(\mu-\mu_c)/T_c$, $t=(T-T_c)/T_c$ where the scaling law: $\langle N \rangle \sim M^{\tilde{q}}$ with $\tilde{q} \in (0.75,1)$ holds. The blue line is the line for which $\tilde{q}=0.96$. The freeze-out states for central A+A collisions in NA49 experiment at maximum SPS energy according to \cite{Becattini2006} are given by the stars: A=Pb (black), A=Si (blue), A=C (orange).
(b) The line describes the possible pairs $(T_c,\mu_c)$ compatible with the finite-size analysis in the current Letter. The red patterned region is determined employing the different Lattice QCD results for the critical temperature $T_c$ \cite{Katz2004,Gupta2008}.}
\label{fig:pdcr}
\end{figure}   

To illustrate in more detail how our strategy leading to the critical region in Fig.~1a works in practice, we plot in Fig.~2 in double logarithmic scale, the mean value $\langle N \rangle$ versus $M$ for various values of $\hat{m}$ and $\zeta$. The plot focuses on the relevant region $310 \leq M \leq 700$. To facilitate the comparison we have scaled all moments to a common value at $M=310$ using as reference  the black line ($\hat{m}=0$ and $\zeta=1$). In Fig.~2a we show $\langle N \rangle$ for $\hat{m}=0,~0.07,~0.15$ and $\zeta=1$.  We observe that although $\hat{m}$ increases by a factor of two or equivalently the reduced temperature $t=\frac{T-T_c}{T_c}=\hat{m}^{3/2}$ by a factor of three (compare blue and red lines in Fig.~2a), the corresponding slope change is relatively small. This slow change of the exponent $\tilde{q}$ with increasing $\hat{m}$ explains why the critical region in the $t$ (or $\hat{m}$)-direction is wide. 
On the other hand, assuming $\hat{m}=0$ and varying $\zeta$ across the real axis, we observe that the critical scaling goes over to the conventional behaviour $\langle N \rangle \sim M$ quite rapidly. This behaviour is in accordance with the plot of the critical region in Fig.~1a and it is clearly demonstrated in Fig.~2b where we plot in double logarithmic scale $\langle N \rangle$ versus $M$ for three different $\zeta$-values ($1$, $1.01$, $1.02$) using $\hat{m}=0$ ($T=T_c$). 
Notice that, in Fig.~2a, the slope decreases as we depart from the critical (black) line, going over to the mean field behaviour, while in Fig.~2b it increases, going over to the conventional behaviour mentioned above. Furthermore, it is worth also to mention that the value of $\langle N \rangle$, obtained through Eq.~(\ref{eq:4}), is a prediction for the total mean baryon-number multiplicity in the critical freeze-out state, since there are no free parameters in the calculation. This information is shown by the black line in Fig.~2a (or Fig.~2b) which gives the mean baryon number as a function of the system's size $M$. 

\begin{figure}[tbp]
\centering
\includegraphics[width=0.5\textwidth]{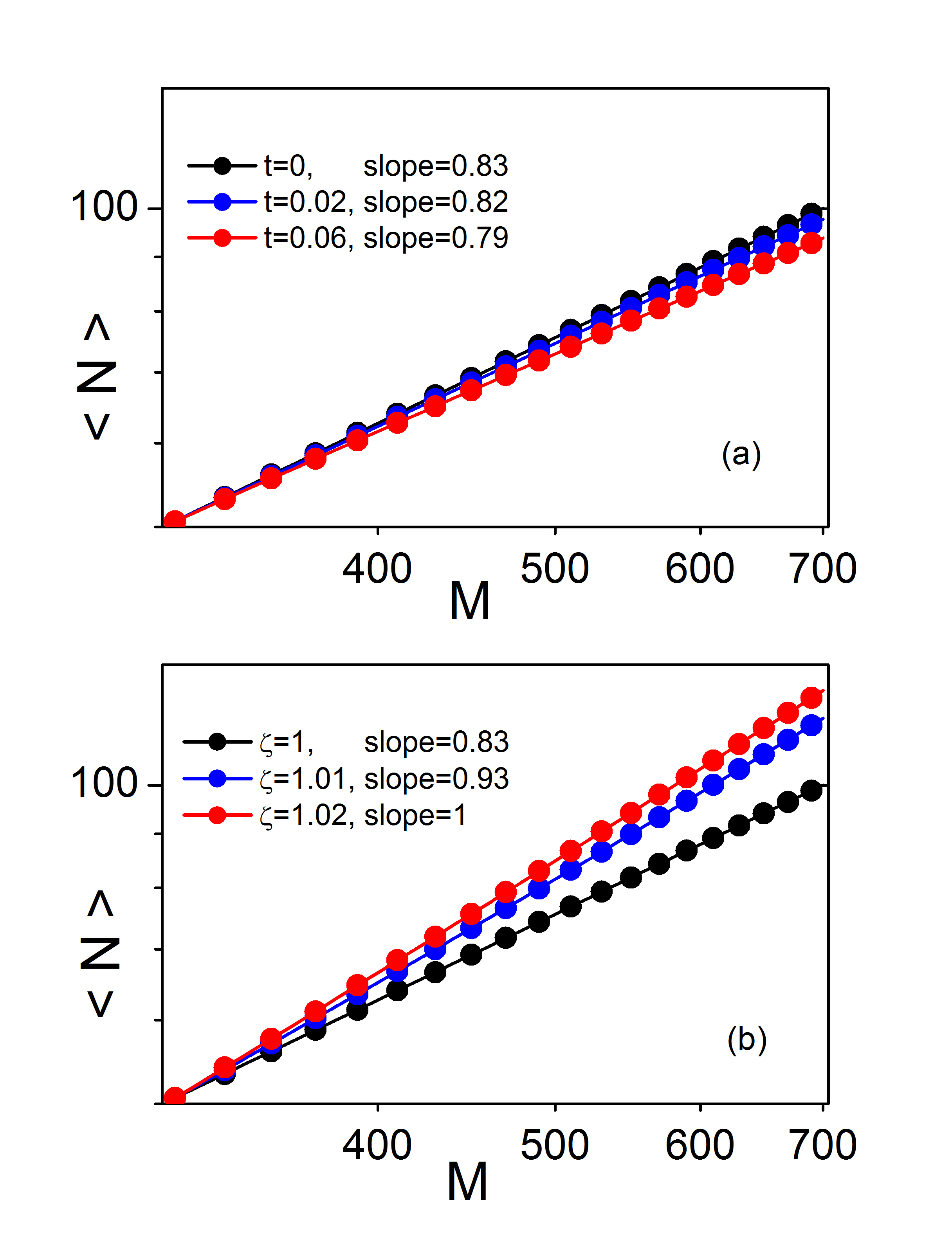}
\caption{(a) The first moment of the baryon number density $\langle N \rangle$ as a function of $M$ for $t=0,~0.02,~0.06$  (corresponding to $\hat{m}=0,~0.07,~0.15$) and $\zeta=\zeta_c=1$. (b) The mean value  $\langle N \rangle$ as a function of $M$ for $T=T_c$ ($\hat{m}=0$) and $\zeta=1,~1.01,~1.02$.}
\label{fig:ly}
\end{figure}   

Finally, it is important to demonstrate that the effective action (\ref{eq:1}) is fully consistent with the critical properties of the 3d-Ising universality class. Therefore, we extend our finite-size analysis to the treatment of baryon-number susceptibility. From fluctuation-dissipation theorem we have:
\begin{equation}
\chi=\frac{1}{V} \langle (\delta N)^2 \rangle~~~;~~~\langle (\delta N)^2 \rangle = \zeta \frac{\partial}{\partial \zeta} \left( \zeta \frac{\partial \log \mathcal{Z}}{\partial \zeta} \right)
\label{eq:7}
\end{equation}
At the critical chemical potential ($\zeta_c=1$) we expect a peak of $\chi$ at $T=T_c$ ($\hat{m}=0$) for large but finite $M$ obeying the finite-size scaling relation
\begin{equation}
\chi(T_c) \sim V^{\gamma/\nu d} \sim (k_B T_c)^3 M^{2 q -1}
\label{eq:8}
\end{equation}
with $2 q -1 =\frac{\gamma}{\nu d}$ ($=2/3$ for the 3d-Ising universality class), as well as a power-law dependence on temperature
\begin{equation}
\chi(T) \sim (T-T_c)^{-\gamma}~~~~~;~~~~~M \to \infty
\label{eq:9}
\end{equation}
close to the CEP (with critical exponent $\gamma \approx 4/3$ for the infinite system). In Fig.~3, plotting $\chi(T)/M^{2 q - 1}$, calculated numerically through the partition function (\ref{eq:3}), for various values of $M$, we clearly show the validity of both scaling laws in Eqs.~(\ref{eq:8},\ref{eq:9}).

\begin{figure}[tbp]
\centering
\includegraphics[width=0.5\textwidth]{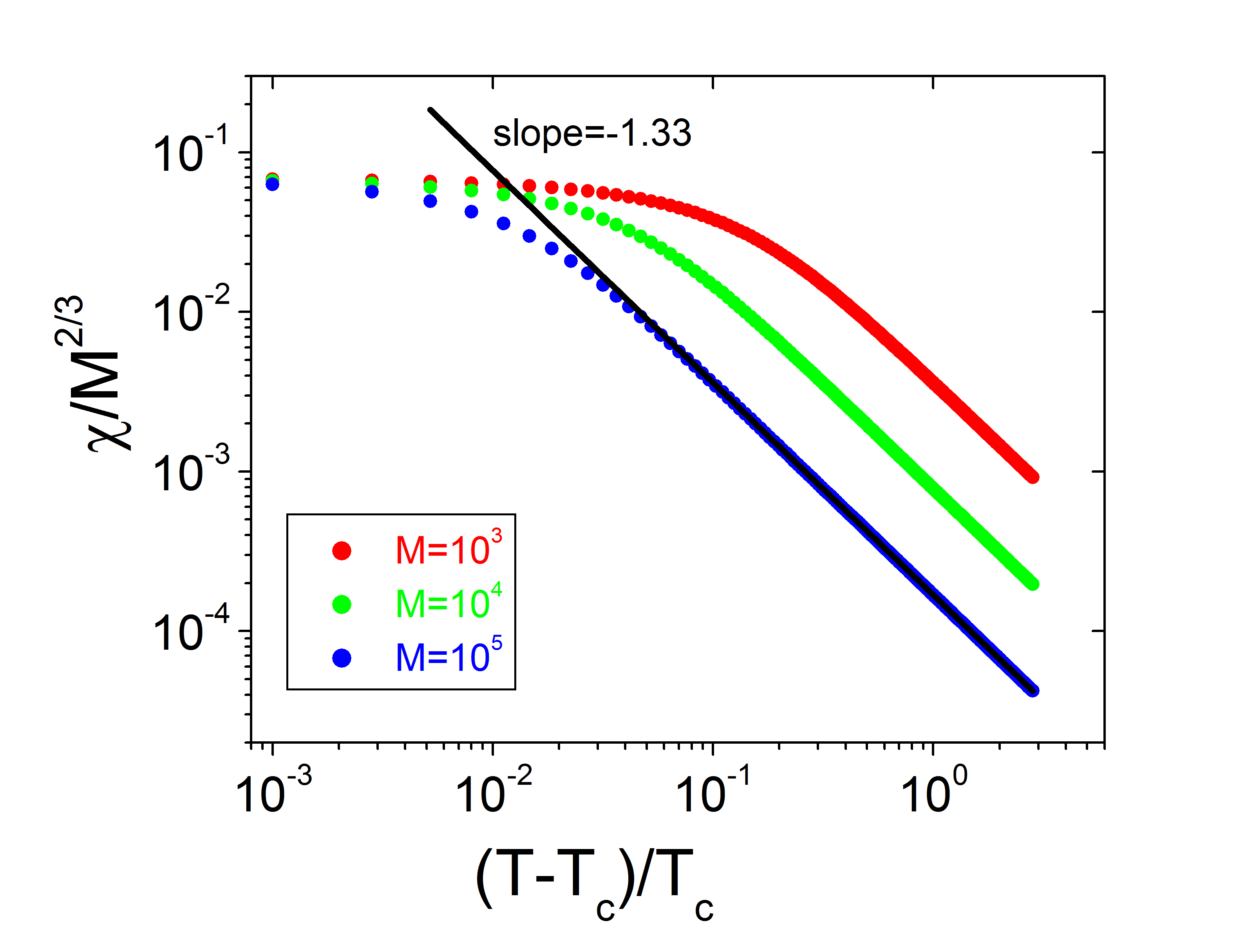}
\caption{The baryon number susceptibility scaled by $M^{2/3}$ versus the reduced temperature $(T-T_c)/T_c$ plotted in double logarithmic scale for 3 different values of the size $M$: $10^3$ (red circles), $10^4$ (green circles) and $10^5$ (blue circles). The slope in the linear fit determines the critical exponent $\gamma$. All sets converge to the same value for the scaled susceptibility for $T \to T_c$ verifying the finite-size scaling relation (\ref{eq:8}). }
\label{fig:sc}
\end{figure}   

Thus the effective action (\ref{eq:1}) captures correctly the 3d-Ising critical behaviour, supporting strongly the validity of the finite-size scaling analysis in the present Letter.

Concluding remarks are now in order. Based on the effective action proposed in \cite{Tsypin1994} for the 3d-Ising system we have demonstrated that the critical region of the QCD CEP is wide along the temperature direction and very narrow along the chemical potential axis. Using published results on intermittency analysis of proton density fluctuations in SPS NA49 experiment \cite{Anticic2015} and Lattice QCD estimates of the critical temperature it is possible to give a prediction for the location of the QCD CEP $(\mu_c,T_c) \approx (256,150)$ MeV, as shown in Fig.~1b. 

From the analysis, above, one may draw conclusions about the most promising, crucial measurements in the experiments, currently in progress, at CERN and BNL (SPS-NA61, RHIC-BES). It is suggestive from the constraints of the critical region in Fig.~1 that
2d intermittency of net-proton density in transverse momentum space (in the central rapidity region) combined with chemical freeze-out measurements may capture the systems, for different energies ($\sqrt{s_{NN}}$) and size of nuclei (A), which freeze out very close to the critical point. To this end we consider two classes of experiments with heavy (I) and medium or small size (II) nuclei:

\noindent
{\bf I.} \underline{Pb+Pb, Au+Au}: The crucial energy range for these processes in the experiments at CERN (Pb+Pb) and BNL (Au+Au), compatible with the requirements of the critical region (Fig.~1) is 
\begin{itemize}
\item Pb+Pb at $12.3 ~\textrm{GeV} < \sqrt{s_{NN}} < 17.2~ \textrm{GeV}$ (SPS-NA61) corresponding to lab-energies $80~\textrm{AGeV} < E_{lab} < 158~\textrm{AGeV}$ and
\item Au+Au at $14.5 ~\textrm{GeV} < \sqrt{s_{NN}} < 19.6~ \textrm{GeV}$ (RHIC-BES)
\end{itemize}

\noindent
{\bf II.} \underline{Be+Be, Ar+Sc, Xe+La}: The crucial measurements in these processes, regarding 2d intermittency, are in progress at the experiment SPS-NA61, at the highest SPS energy $\sqrt{s_{NN}}=17.2$ GeV. To complete the picture, however, a detailed study of chemical freeze-out in these collisions is also needed.

\end{document}